\documentclass[%
floatfix,
twocolumn,
showpacs,
superscriptaddress,
 amsmath,amssymb,
 prb
]{revtex4-1}

\usepackage[dvipdfmx]{graphicx}
\usepackage{dcolumn}
\usepackage{bm}

\newcommand{\subm}[1]{_{\mathrm {#1}}}

\newcommand{\Tc}{\ensuremath{T\subm{c}}}

\newcommand{\Hcc}{\ensuremath{H\subm{c2}}}
\newcommand{\Hdc}{\ensuremath{H\subm{dc}}}

\newcommand{\sro}{Sr$_2$RuO$_4$}

\begin{document}

\preprint{APS/123-QED}

\title{Quenched metastable vortex states in Sr$_{2}$RuO$_{4}$}

\author{D. Shibata}
\affiliation{%
 Department of Physics, Graduate School of Science, Kyoto University, Kyoto 606-8502, Japan
}%

\author{H. Tanaka}
\affiliation{%
 Faculty of Science, Kyoto University, Kyoto 606-8502, Japan
}%

\author{S. Yonezawa}
\affiliation{%
 Department of Physics, Graduate School of Science, Kyoto University, Kyoto 606-8502, Japan
}%

\author{T. Nojima}
\affiliation{
 Institute for Materials Research, Tohoku University, Sendai 980-8577, Japan
}
\author{Y. Maeno}%
\affiliation{%
 Department of Physics, Graduate School of Science, Kyoto University, Kyoto 606-8502, Japan
}%


\date{\today}

\begin{abstract}
\sro, a leading-candidate spin-triplet superconductor and a highly anisotropic quasi-two-dimensional type-II superconductor,
 provides unique opportunity to study unconventional as well as conventional vortex phases.
To investigate its vortex-matter phases, we studied the ac susceptibility of \sro\ for fields parallel to the RuO$_2$ plane by adapting two different thermal processes:
In addition to the ordinary field sweep~(FS) process, 
we newly employed the ``each-point field cooling~(EPFC)'' process, in which the superconductivity is once thermally destroyed before collecting the data.
We find that the ac susceptibility signal substantially changes with the EPFC process. 
This result indicates that we succeed in inducing new metastable vortex states by the EPFC process. 
We also find a new field scale $H^{\**1}$, below which the FS and EPFC processes provide the same value of the ac susceptibility. 
This new field scale suggests a liquid-like vortex state in the low-field region.
\end{abstract}

\maketitle


\section{\label{sec:level1}introduction}
In a large part of the field-temperature superconducting phase diagram of a type-II superconductor below its critical temperature $\Tc$,
 magnetic quantum vortices penetrating the superconductor exhibit a number of vortex-matter phases depending on strengths of vortex-vortex interaction, thermal fluctuation, pinning energies by lattice defects, etc.
Although such vortex-matter phases have been extensively studied in high-\Tc\ cuprate superconductors,~\cite{Blatter1994,Fisher1991,Nelson1989,Nelson1988,Fisher1989,Zeldov1995}
superconductors with lower $\Tc$~(e.g. NbSe$_{2}$, CeRu$_{2}$, etc.) also exhibit vortex phase transitions.~\cite{Banerjee2001,Tomy2002,Nattermann1998,Ravikumar1998}
In such low-$\Tc$ superconductors, vortices at low fields form a quasi-long-range-ordered lattice, which is called as the vortex Bragg glass~(VBG).
This is because the arrangement of vortices is mostly governed by the repulsive vortex-vortex interaction.
As the magnetic field is increased, vortex lattice  becomes softer, because the vortex-vortex interaction becomes weaker.
Then, near the upper critical field $\Hcc$, other interactions can become dominant.
As a result, the VBG state changes into the vortex glass~(VG) state, 
in which vortices form a glassy structure without long-range ordering due to a dominance of vortex pinning. 
As the field is further increased, thermal fluctuation then becomes dominant and the VG melts into the vortex liquid~(VL), where vortices can move individually.

It is now widely known that the measurement of the ac susceptibility $\chi\subm{ac}\equiv\chi'-i\chi''$ is a powerful technique to explore the vortex matter phases.
One example is the peak effect,~\cite{Banerjee2001} which is the occurrence of an anomalous maximum in the shielding signal $|\chi'|$ near $\Hcc$.
In weakly pinned superconductors, the onset of the peak effect corresponds to the VBG-VG transition and the peak top corresponds to VG-VL transition. 
Another example is hysteretic behavior in $\chi\subm{ac}$.
It is often observed that the field cooling~(FC) process and the zero field cooling~(ZFC) process lead to different values of $\chi\subm{ac}$.
Such differences originate from hysteretic vortex configurations mainly caused by vortex pinning.
In typical cases, an ordered vortex lattice state and a disordered vortex amorphous/glass-like state are achieved by the ZFC and FC processes, respectively.~\cite{Ling2001}\\

 \sro, a layered perovskite superconductor with $\Tc=1.5$ K, has been extensively studied due to its unconventional pairing state.~\cite{Maeno1994,Kallin2012.RepProgPhys.75.042501,Maeno2012.JPhysSocJpn.81.011009,Mackenzie2003RMP}
 Spin susceptibility measurements with the nuclear magnetic resonance~(NMR) and with the polarized neutron scattering indicate the spin part of the Cooper pair is in the spin-triplet state.~\cite{Ishida1998.Nature.396.658,Ishida.unpublished,Miyake2014.JPhysSocJpn.83.053701,Duffy2000.PhysRevLett.85.5412}
 Muon-spin rotation and optical Kerr effect studies have revealed that the superconducting state of \sro\ is of chiral-$p$-wave,
 in which two degenerate order parameters form a complex linear combination, breaking the time reversal symmetry.~\cite{Luke1998.Nature.394.558,Xia2006.PhysRevLett.97.167002}
 Recently, a non-trivial topological nature of the chiral-$p$-wave superconducting wave function has been attracting wide attention.~\cite{Maeno2012.JPhysSocJpn.81.011009}
 Such a chiral-$p$-wave spin-triplet superconductor has spin and orbital degrees of freedom in its superconducting order parameter. 
 Thus, it is expected that magnetic field affects such degrees of freedom and leads to emergence of new superconducting phases.~\cite{Agterberg1998.PhysRevLett.80.5184,Udagawa2005,Kaur2005.PhysRevB.72.144528,Yanase2014.JPhysSocJpn.83.061019}
 Indeed, such multiple phases have been observed in UPt$_{3}$.~\cite{Joynt2002} In case of \sro, a previous specific-heat study reported possible existence of new phases in the vicinity of $\Hcc$.~\cite{Deguchi2002}
 However, more recent studies with a smaller crystal did not reproduce the result.~\cite{Yonezawa2013.PhysRevLett.110.077003,Yonezawa2014}
 Thus, it is still an open question whether the superconducting multiphase exists in \sro\ or not.

 \sro\ has another interesting aspect as a highly anisotropic quasi-two-dimensional superconductor, reflecting its layered crystal structure.
 The anisotropy in the upper critical field $\Hcc$, $\varGamma_{H} = H_{\mathrm{c2}\parallel ab}/H_{\mathrm{c2}\parallel c}$, is 20 for $T\rightarrow 0$.~\cite{Maeno2012.JPhysSocJpn.81.011009}
 On the other hand, recent studies indicate that the intrinsic superconducting anisotropy $\varGamma_{\mathrm{I}} = \xi_{ab}/\xi_{c}$ is as large as 60.\cite{Rastovski2013.PhysRevLett.111.087003,Kittaka2014}
 This large anisotropy may lead to interesting vortex phase formation in this material.
 In particular for $H\parallel ab$, competition among various length and energy scales can lead to non-trivial vortex phases.
 For example, realization of vortex liquid crystals in an anisotropic type-II superconductor has been theoretically proposed~\cite{Carlson2003}.
 Note that, however, its coherence length along the $c$ axis is estimated to be $\xi_{c}(0)\sim \xi_{a}(0)/60 \sim 13$~\AA, being still larger than the interlayer spacing $c/2\sim 6$~\AA.
 Thus, interlayer coherence should be maintained in \sro, in clear contrast to high-$\Tc$ cuprates, whose interlayer coherence can be vanishingly small.

Considering these situations, investigation of the vortex phase diagram in \sro\ is quite interesting and important. 
Firstly, to distinguish between the ordinary vortex-matter phases and unconventional superconducting multiphases originating from the spin-triplet order parameter, it is important to understand the vortex phase diagram in detail.
Secondly, this oxide provides an unique opportunity to study vortex phases in highly anisotropic low-$\Tc$ superconductivity (i.e. with relatively large coherence length).
Previously, the small-angle neutron scattering~(SANS) measurement revealed vortex lattices in some regions of the $H-T$ phase diagram both for $H\parallel a$ and $H\parallel c$.~\cite{Riseman1998,Rastovski2013.PhysRevLett.111.087003}
However, details of the vortex phase diagram has not been explored.
In addition, although $\chi\subm{ac}$ of \sro\ has been reported,~\cite{Yoshida1996,Yaguchi2002.PhysRevB.66.214514,Kittaka2009.PhysRevB.80.174514} effects of different thermal/field treatments have not been investigated.


In this paper, we report $\chi\subm{ac}$ of \sro\ for $H\parallel ab$ measured with a newly developed thermal/field process as well as with a conventional field-sweep process.
With the new process, we succeed in systematically inducing metastable vortex states. 
By comparing $\chi\subm{ac}$ results with different thermal treatments, we obtain a vortex phase diagram with a new phase boundary at low fields.\\

\section{\label{sec:level2}Experimental}


We used single crystalline \sro \ grown by a floating-zone method.~\cite{Mao2000.MaterResBull.35.1813} 
The sample used in this study has the size of $1.5 \times 6.0 \times 0.1~$mm$^{3}$, and
was cut from the crystal boule used for the SANS measurements.~\cite{Rastovski2013.PhysRevLett.111.087003} 
Zero-field $\chi\subm{ac}$ measurements revealed the transition at $\Tc=1.45$~K, which is defined as the mid-point temperature of the real part $\chi'$. 
The directions of the tetragonal crystalline axes of the sample were determined from X-ray Laue pictures.
After cut, we glued two strain gauges with a resistance of 120~$\Omega$ (Kyowa Dengyo, KFRS-02-120-C1-13 L1M3R) as heaters onto the $ab$-surfaces of the sample directly.
We heat-treated the sample at up to 150${}^\circ$C for several hours in order to glue heaters.
This heat treatment may have served as a gentle annealing process.


 The sample together with the heaters was glued to a sapphire rod with varnish (GE7031) and placed in a mutual-inductance coil,
consisting of a counter-wound pick-up coil with 3300 turns and an excitation coil with 860 turns.
A string of gold wire~($\phi = 25~\mathrm{\mu m}$) is also attached between the sample and the thermal bath to achieve faster thermal equilibrium.
Measurements of $\chi\subm{ac}$ were performed with an ac field of 0.66~$\mu$T-rms at the frequency of 3011~Hz.
The direction of the ac field is within the $ab$ plane and about 10 degrees away from the [100] axis. 
The configuration of the sample assembly and magnetic field directions is schematically shown in Fig.~1(a).
The sample assembly was cooled down to below 0.1 K with a $^{3}$He-$^{4}$He dilution refrigerator (Oxford Instruments, Kelvinox-25). 
The dc magnetic field $H\subm{dc}$ was applied using a vector magnet system~\cite{Deguchi2004RSI}. Based on the strong anisotropy of $\Hcc$ of \sro, 
the field directions were determined with accuracies within $0.01^\circ$ with respect to the $ab$ plane and within $1^\circ$ with respect to the direction within the plane.

\begin{figure}[htbp]
\includegraphics[width=3.3in]{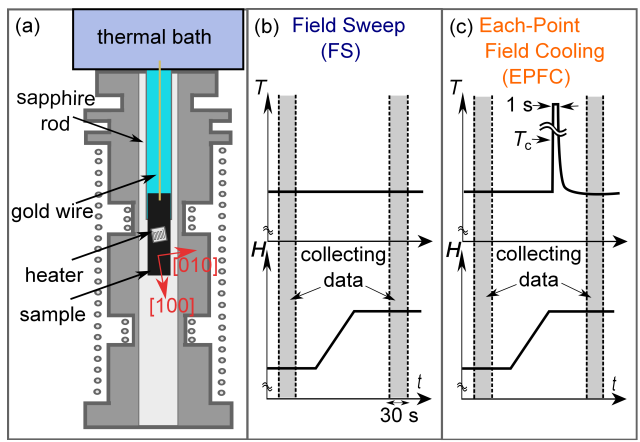}
  \caption{
  (a) Schematic of the sample mounted in an ac susceptmeter coils. The sample, shown in black, is placed in one part of the pick-up coil.
  Two different thermal processes, (b) field sweep (FS) process and (c) each-point field cooling (EPFC) process, were used in this study.
  }
  \label{fig01}
\end{figure}


In this paper, we measured $\chi\subm{ac}$ in two different thermal processes, as depicted in Figs.~1(b) and (c). 
The first process is an ordinary field sweep~(FS): the sample was cooled down to each target temperature in zero field and, kept at that temperature,
the data was collected at each field on a field up-sweep sequence, followed by a field down-sweep sequence.
The second process is the ``each-point field cooling~(EPFC)''.
In this process, the sample at each field $H\subm{dc}$ was once quickly heated up to above $\Tc (H\subm{dc})$ using the sample heaters and
cooled back to the target temperature before collecting the data as described in Fig.~1(c).
We confirmed by the $\chi\subm{ac}$ signal that the sample is indeed heated up to above $\Tc (H\subm{dc})$ to become the normal state.
After collecting the data, $H\subm{dc}$ was changed before the next sequence of the heating, cooling, and data collection.
As shown below, metastable vortex states can be induced with this latter process. 
We comment here that the EPFC process is similar to that adapted in Ref~\citenum{Ravikumar1998}.
However capability of much faster heating and cooling of the sample in our study enables us to construct vortex phase diagrams efficiently and precisely.
Although we measured both the real and imaginary parts of $\chi\subm{ac}$, the imaginary part was too small.
Thus, in this paper, we only discuss the real part.
In order to subtract $\chi'$ contribution from normal state and background, we adopt $\chi'\subm{SC}\equiv\chi'(T,H)-\chi'($2.0~K,$H)$ as already done in Ref.\citenum{Kittaka2009.PhysRevB.80.174514}.
We furthermore scaled $\chi'\subm{SC}$ so that $\chi'\subm{SC} = -1$ at $H\subm{dc} = 0$ and $T = 0.1$~K.

\section{\label{sec:level3}Results}

In Fig.~2(a), we present $\chi'\subm{SC}(H\subm{dc})$ obtained with the FS and EPFC processes for $H\subm{dc}\parallel [110]$ at several temperatures.
First we focus on results of the ordinary FS process.
The $\chi'\subm{SC}(H\subm{dc})$ curve of the FS process has a characteristic peak/dip structure near $\Hcc$.
This behavior resembles the peak effect observed in many ordinary type-II superconductors,~\cite{Tomy2002,Nattermann1998,Banerjee2001}
 and is consistent with previous reports on \sro.~\cite{Yoshida1996,Mao2000.PhysRevLett.84.991,Yaguchi2002.PhysRevB.66.214514,Kittaka2009.PhysRevB.80.174514}
The origin of this behavior will be discussed later.
Note that the curves contain data of field-up and down sweeps; We didn't observe any hysteresis between field-up and down sweeps except for near $\Hcc$ at low temperatures where the superconducting transition is of first order~\cite{Yonezawa2013.PhysRevLett.110.077003,Yanase2014.JPhysSocJpn.83.061019}.
This absence of hysteresis in the FS branch in the superconducting state implies that the pinning effects are rather weak in this material 
so that one cannot achieve metastable vortex phases with the ordinary FS process.

Next, we focus on the $\chi'\subm{SC}$ data of the EPFC process for $H\subm{dc}\parallel[110]$ in Fig.~2(a).
Interestingly, the $\chi'\subm{SC}(H\subm{dc})$ curves for the EPFC process are very different from those for the FS process.
Thus, we succeeded in inducing new vortex states by the EPFC process.
The shielding signal $|\chi'\subm{SC}(H\subm{dc})|$ in the FS branch is smaller than that in the EPFC branch in all investigated field and temperature conditions.
This behavior indicates that the vortices in the EPFC branch are harder to move than those in the FS branch. 

\begin{figure}[htbp]
\includegraphics[keepaspectratio,width=3.3in]{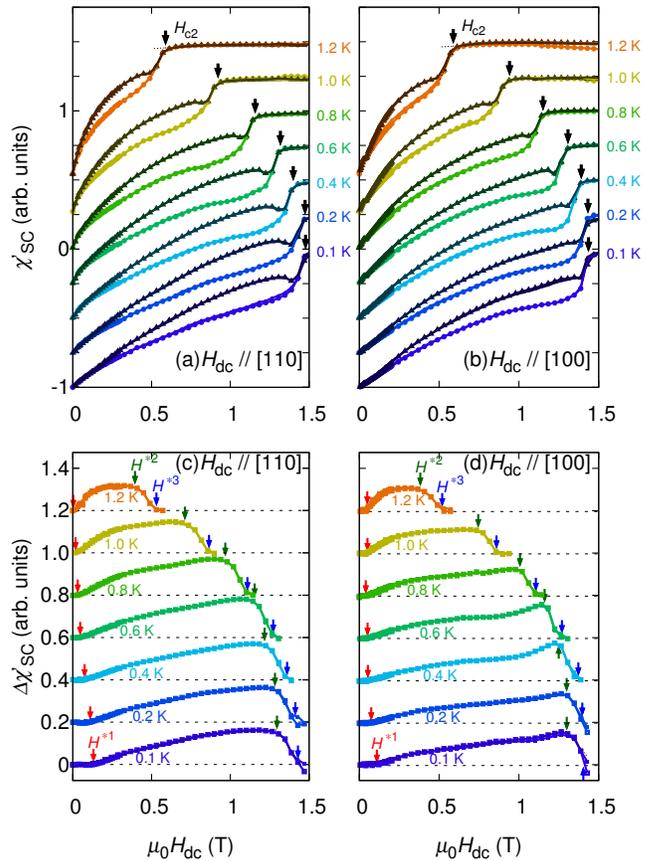}
  \caption{
  (a)-(b) In-plane dc field dependence of the real part of the ac susceptibility $\chi\subm{SC}'$ of \sro\ at various temperatures 
  for $H\subm{dc} \parallel $[110] and $H\subm{dc} \parallel $[100].
  The triangles indicate data measured in FS processes and the circles with EPFC processes.
  Each curve is shifted vertically by 0.25 for clarity.
  (c)-(d) Dependence on dc field of $\Delta\chi'\subm{SC}\equiv\chi'\subm{SC}($FS$)-\chi'^{\mathrm{ave}}\subm{SC}($EPFC$)$. 
  Each curve is shifted vertically by 0.2 for clarity.
  The arrows indicate $H^{\**1}$, $H^{\**2}$, and $H^{\**3}$ characterizing the field dependence of $\Delta\chi'\subm{SC}$ (see text).
  All panels contain data for field up sweep (open symbols) and down sweep (closed symbols).
  }
\label{fig02}
\end{figure}

In order to investigate the stability of the EPFC-induced state, we tried other field/thermal processes.
Firstly, we applied a small dc-field cycling with the amplitude $\delta\Hdc$ to the EPFC branch just before collecting data as schematically explained in the inset of Fig.~3(a). 
As shown in Fig~3(a), the EPFC branch changes toward the FS branch with increasing $\delta\Hdc$.
A field cycling of as small as 0.001~T is sufficient for the EPFC branch to merge into the FS branch.
Larger field-cycling amplitude does not change the FS branch to other branches.
Secondly, we performed ordinary FS after a EPFC process at $\mu_{0}H\subm{dc} = 1.263$~T as presented in Fig.~3(b).
We found that $\chi'\subm{SC}$ rapidly changes once the FS process started.
Then $\chi'\subm{SC}$ becomes constant at a value very close to that of the FS branch when we swept by more than 0.001~T.
These two results are consistent each other and indicate that the vortex state in the EPFC branch is metastable.
Our observation also agrees with the general tendencies that the field-cooled vortex states are metastable,
and that magnetic field cycling recovers more stable vortex states.~\cite{Pasquini2008}

To compare the FS and EPFC branches in more detail, we evaluate the difference $\Delta\chi'\subm{SC}\equiv\chi'\subm{SC}($FS$)-\chi'^{\mathrm{ave}}\subm{SC}($EPFC$)$,
where $\chi'^{\mathrm{ave}}\subm{SC}($EPFC$)$ is the average of $\chi'\subm{SC}($EPFC$)$ in up and down sweeps.
Note that $\Delta\chi'\subm{SC}$ provides a measure of how sensitive the vortex configuration is against field/thermal processes.
From the $\Delta\chi'\subm{SC}$ data in Fig.~2(c), we can identify several field regions with different responses against field/thermal processes.
Firstly, in the lowest field region below a field that we denote by $H^{\**1}$, $\Delta\chi'\subm{SC}$ is almost zero, indicating that the FS and EPFC branches exhibit the same value.
Thus, the vortex configurations achieved by the FS and EPFC processes are nearly the same for fields below $H^{\**1}$.
We emphasize that this field scale $H^{\**1}$ in \sro\ has not been reported before to our knowledge and is newly revealed owing to the EPFC process employed in the present work.
Secondly, above $H^{\**1}$, $\Delta\chi'\subm{SC}$ becomes larger with increasing field. On further increase of the field, $\Delta\chi'\subm{SC}$ starts to decrease beyond an onset field $H^{\**2}$.
In this region, the FS branch shows a peak-effect-like feature as already explained.
Above a field which we denote by $H^{\**3}$, $\Delta\chi'\subm{SC}$ becomes zero again and both processes yield the same vortex state, probably due to small pinning effect.\\


\begin{figure}[Htbp]
\begin{center}
\includegraphics[keepaspectratio,scale=0.8]{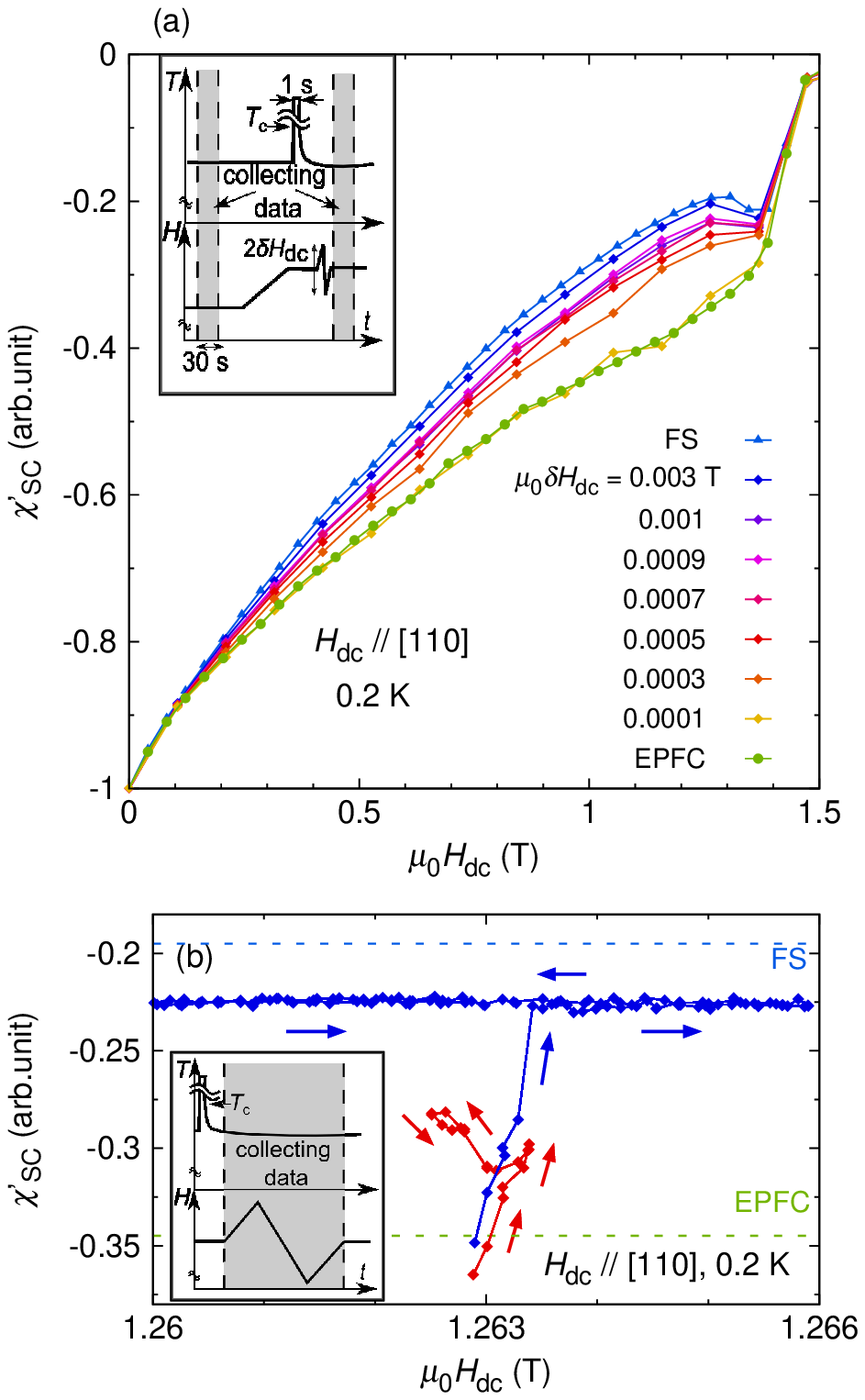}
  \caption{
  Stability of the EPFC branch against a small cycling of the dc field.
  (a) Changes of the EPFC state toward the FS state after a small dc-field cycling $\delta\Hdc$ to a state on the EPFC branch.
  (b) Results of FS measurements after a EPFC process at $\mu_0H\subm{dc} =$ 1.263~T,
  revealing detailed evolution of the EPFC branch toward the FS branch by a field cycling.
  The directions of the FS are indicated with arrows.
  The values of $\chi'\subm{SC}$ in the FS and EPFC processes at 1.265~T are indicated with the broken lines.
  The insets explain field/thermal treatments in these measurements.
  }
\label{fig03}
\end{center}
\end{figure}

 From these data, we construct the vortex phase diagram of \sro\ for $\Hdc\parallel[110]$ as presented in Fig.~4(a). 
 For the phase diagram, we adopt the following definitions for the field scales:
We define $H^{\**1}$ as the field where the linear extrapolation in the region $\Hdc>H^{\**1}$ intersects the $\Delta\chi'\subm{SC} = 0$ line,
$H^{\**2}$ as the field where the linear extrapolations of $\Delta\chi'\subm{SC}$ increasing and decreasing regions intersects each other,
and $H^{\**3}$ as the field where the linear extrapolation of the region $\Hdc<H^{\**3}$ intersects the $\Delta\chi'\subm{SC} = 0$ line, respectively.
These three characteristic fields decrease with increasing temperature but remain finite up to the zero-field $\Tc$.
The narrow region between $\Hcc$ and $H^{\**3}$ become a little wider at low temperature.
The separation between $\Hcc$ and $H^{\**2}$ is nearly independent of temperature below about 1~K.
We note that the metastable state can be induced by the EPFC process in the wide region of the superconducting state surrounded by the $H^{\**1}$ and $H^{\**3}$ curves.

%

\begin{figure}[Htbp]
\includegraphics[keepaspectratio,scale=0.8]{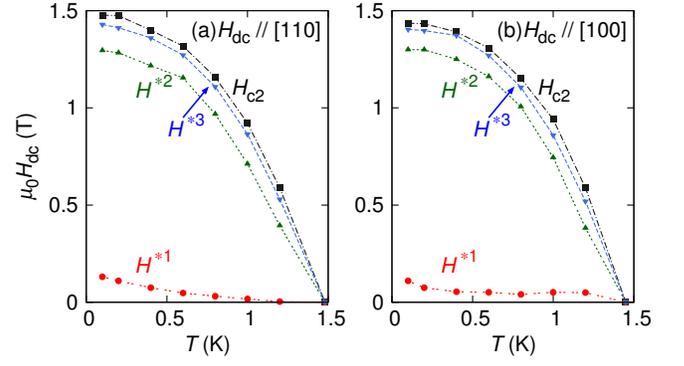}
 \caption{
  (a)-(b) Field-temperature phase diagrams of \sro\ for the dc field along the [110] and [100] directions.
  The black squares indicate $\Hcc$.
  The red circles, green triangles, and blue triangles indicate $H^{\**1}$, $H^{\**2}$, and $H^{\**3}$, respectively, derived from the comparison between FS and EPFC branches (see text).
 }
\label{fig04}
\end{figure}


We performed similar measurements and analyses for $\Hdc\parallel [100]$ to investigate the in-plane anisotropy.
As shown in Fig.~2(b), $\chi'\subm{SC}$ curves for $\Hdc\parallel [100]$ are similar to those for $\Hdc\parallel [110]$:
The FS curves exhibit peak/dip structure near $\Hcc$ and the shielding signal in the FS branches is smaller than that in the EPFC branches.
However there are some differences. 
For example, although the $\chi'\subm{SC}$ curves for $\Hdc\parallel [110]$ at 0.1~K and 0.2~K show a clear peak/dip structure,
such structure is rather vague in $\chi'\subm{SC}(\Hdc)$ for $\Hdc\parallel [100]$.
In addition, the temperature dependence of $H^{\**1}$ appears different:
$H^{\**1}$ for $\Hdc\parallel [110]$ monotonically decreases as the temperature is increased toward $\Tc$, whereas $H^{\**1}$ for $\Hdc\parallel [100]$ exhibits a plateau in the range 0.5~K $< T <$ 1.2~K.
As a result, above 0.8~K, $H^{\**1}$ for [100] is much larger than that for [110].
We comment here that this difference may be attributed to the difference in the vortex-lattice stability due to the in-plane anisotropy in the superconducting order parameter.
Although previous $\chi\subm{ac}$ study reported additional peak structure near $\Hcc$ only in the [110] direction,~\cite{Yaguchi2002.PhysRevB.66.214514}
 we did not observe such additional peaks in this study possibly because the number of data points was not enough.



\section{\label{sec:level4}Discussion}

First, we briefly summarize our experimental observations. 
The ordinary FS branch of \sro\ exhibits peak/dip structure in $\chi'\subm{SC}$ in the field region $H^{\**2} < \Hdc < H^{\**3}$,
 similar to the peak effect in ordinary type-II superconductors such as NbSe$\subm{2}$.~\cite{Tomy2002}
The peak effect is attributed to changes in the vortex phases: the low-field onset of the peak effect correspond to 
the transition between the ordered VBG state in low fields and the disordered VG state, and the peak top corresponds to the transition between the VG and VL states.
 Therefore, in the ordinary type-II superconductor, $H^{\**2}$ and $H^{\**3}$ should correspond to VBG-VG and VG-VL transition lines.
Similar situation is probably realized in \sro, although consideration of the first-order superconducting transition is needed as described below. 
Indeed, previous SANS experiments~\cite{Riseman1998,Rastovski2013.PhysRevLett.111.087003} have revealed that the vortices exhibit clear Bragg reflections in wide regions of the phase diagram for both $\Hdc\parallel a$ and $\Hdc\parallel c$,
 with thermal/field processes corresponding to our FS process.
These results indicate that the bulk VBG state is formed in such regions.
We achieved new metastable vortex phases by the EPFC process in the region $H^{\**1}<\Hdc<H^{\**3}$.
By examining the difference between the EPFC and FS branches, we are able to determine the accurate value of the onset $H^{\**2}$ of the peak-effect-like feature up to a high-temperature region,
 where the peak in $\chi'\subm{SC}(\Hdc)$ itself becomes rather vague.
Furthermore, we can get the new characteristic field $H^{\**1}$ signaling disappearance of metastability in the low-field region.


For $H^{\**1} < H < H^{\**2}$, we revealed that vortex states in the EPFC process is metastable and should differ from the VBG state.
There are several possible scenarios explaining the vortex state in the EPFC process.
The first scenario is that a strongly pinned glassy state is induced by the EPFC process.
Such a glassy state has been indeed reported in ordinary type-II superconductors.~\cite{Tomy2002}
The second scenario is that a cleaner lattice is formed after a EPFC process.
This scenario is based on the fact that the inter-vortex distance only depends on field and cooling process in EPFC should not change the distance.
However, to the best of our knowledge, such a cleaner lattice has not been reported in other superconductors.
In the third scenario, a metastable vortex state with vortices pinned at the surface is realized in the EPFC branch.
Such a surface pinning is called the Bean-Livingston surface barrier and has been indeed observed in high-$\Tc$ superconductors and granular superconductors.~\cite{Burlachkov1993,Rakhmanov1992}
The vortices pinned by the surface pinning potential are easily moved by a small field cycling and are rearranged back to a more stable bulk VBG configuration.

Let us now discuss the origin of the state for $H < H^{\**1}$.
Since the field scale $H^{\**1}$ has previously not been known, 
the VBG state in the bulk was expected to occupy the whole region below $H^{\**2}$ down to the lower critical field $H\subm{c1}$, which is approximately 1~mT at $T \sim 0$.~\cite{Sudershan1997}
Our observation of $H^{\**1}$, however, forces us to reconsider this naive expectation.
In this region, due to a large inter-vortex distance, the vortex-vortex interaction may become too weak to sustain a stable VBG state in the bulk.
For such a region, some theories predict existence of liquid-like vortex state near $H\subm{c1}$ originating from such vanishingly small vortex interactions.~\cite{Nelson1988,Nelson1989,Blatter1994,Fisher1991}
Indeed, such a liquid-like state has been observed in high-$\Tc$ cuprates near $\Tc$.~\cite{Sudershan1997,Hucho2000}
We propose that a similar liquid-like state is realized in \sro\ below $H^{\**1}$.
Note that the first and third scenarios for the stable VBG state for $H^{\**1} < H < H^{\**2}$ are compatible with the formation of the liquid-like state below $H^{\**1}$.
In the first scenario, it is naturally expected that the glassy metastable state becomes difficult to form when a liquid-like state is stable.
In the third scenario, the Bean-Livingston surface barrier is also known to disappear once a liquid-like state is formed.~\cite{}

One apparent issue on this scenario is that the observed $H^{\**1}$ is approximately 100 times larger than $H\subm{c1}$,
whereas ordinarily the low-field liquid phase has been expected only in the vicinity of $H\subm{c1}$.~\cite{footnote01}
To explain why the liquid phase can be realized in such a large field region in \sro , we should consider its large anisotropy.
Assuming a triangular vortex lattice, the inter-vortex distance in an isotropic superconductor is 
\begin{eqnarray}
a\subm{iso} = \sqrt{\frac{2}{\sqrt{3}}\frac{\Phi_{0}}{B}},
\end{eqnarray}
where $\Phi_{0}$ is the flux quantum and $B$ is the magnetic flux density.
Meanwhile, the inter-vortex distances in an anisotropic superconductor are given by 
\begin{eqnarray}
a_{\mathrm{S}} =  \frac{1}{\sqrt{\varGamma}}a\subm{iso}\\
a_{\mathrm{L}} =  \frac{\sqrt{3\varGamma^{2}+1}}{2\sqrt{\varGamma}}a\subm{iso},
\end{eqnarray}
where $a_{\mathrm{S}}$ and $a_{\mathrm{L}}$ are the shorter and longer inter-vortex distances, respectively (Fig.~5); and $\varGamma$ is the superconducting anisotropy.
\begin{figure}[b]
\begin{center}
\includegraphics[width=3.3in,clip]{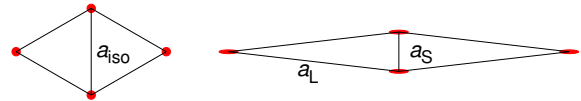}
\caption{Schematic comparison of vortex-lattice configurations for isotropic and anisotropic superconductors. 
The red spots indicate vortex cores.
The expressions for inter-vortex distances $a\subm{iso}$, $a\subm{L}$ and $a\subm{S}$ are given in the text.
 }
\label{fig05}
\end{center}
\end{figure}
It is natural to assume that the liquid state exists when either $a_{\mathrm{L}}$ or $a_{\mathrm{S}}$ becomes longer than a certain length $a\subm{melt}$. 
Then, the melting field for an isotropic superconductor is 
\begin{eqnarray}
B^{\mathrm{iso}}\subm{melt}=\frac{2\Phi_{0}}{\sqrt{3}a^{2}\subm{melt}},
\end{eqnarray}
whereas the melting field for an anisotropic superconductor with $\varGamma \gg 1$ is 
\begin{eqnarray}
B^{\mathrm{ani}}\subm{melt}=\frac{\Phi_{0}(3\varGamma^{2}+1)}{2\sqrt{3}\varGamma a^{2}\subm{melt}}\sim \varGamma\frac{\sqrt{3}\Phi_{0}}{2a^{2}\subm{melt}}.
\end{eqnarray}
Thus, $B^{\mathrm{ani}}\subm{melt}$ is almost $\varGamma$ times larger than $B^{\mathrm{iso}}\subm{melt}$.
For YBa$_2$Cu$_3$O$_{7-\delta}$, $a\subm{melt}$ is expected to be about 6 times larger than the penetration depth $\lambda$.~\cite{footnote02}
Assuming that this relation holds for \sro, we estimate $a\subm{melt} \sim 6\lambda \sim 12,000$~\AA.~\cite{Mackenzie2003RMP}
Thus, using eq.~(5) and $\varGamma = 60$, $B^{\mathrm{ani}}\subm{melt}$ for \sro\ is estimated to be 0.07~T.
Even with this simple estimation, the value of $B^{\mathrm{ani}}\subm{melt}$ semi-quantitatively agrees with the observed $\mu_{0}H^{\**1}\sim 0.1$~T.
We should note here that the vortices may be still ordered along the $c$ direction even below $H^{\**1}$ since $a_{\mathrm{S}}$ is still much smaller than $a\subm{melt}$.
In this sense, the low-field liquid phase may have similarity to liquid crystals, as already proposed in Ref.~\citenum{Carlson2003}.

Before closing the discussion, we comment on the relation between our result and the first-order superconducting transition.
It has been recently revealed that the superconducting to normal transition of \sro\ below 0.8~K under in-plane field is of first-order.~\cite{Yonezawa2013.PhysRevLett.110.077003,Yanase2014.JPhysSocJpn.83.061019}
In ordinary type-II superconductors, continuous suppression of the superconducting order parameter toward zero as $H\rightarrow \Hcc$ is the main source of the formation of the glass and liquid phases near $\Hcc$.
In contrast, the order parameter abruptly disappears from a finite value at the first order transition.
Thus, it is actually surprising that the vortex phase diagrams of \sro\ for $\Hdc\parallel ab$ resemble those of ordinary type-II superconductors.
Note that this sample exhibits weak but clear hysteretic behavior at $\Hcc$, although not clear in the scale of Fig.~2(b).
It is still an open question whether vortex phase transitions can occur near the first-order transition, which should not be accompanied by strong fluctuations.
Further studies with cleaner samples are needed to resolve this interesting issue.


\section{\label{sec:level5}Conclusion}

The EPFC process, in which the sample was once quickly heated up to above $\Tc$ and cooled back to the target temperature before collecting the data, enables us to realize new metastable vortex states.
Precise vortex phase diagrams having a liquid-like state in the low-field region are revealed by comparison between the FS and EPFC branches.
These vortex phase diagrams of \sro\ provide important bases for further studies of searching for superconducting multiphases originating from the anticipated chiral-$p$-wave spin-triplet order parameter.


Finally, we emphasize that the newly employed EPFC process can be adapted not only to ac susceptibility measurements but also to other techniques such as magnetization measurements and neutron diffractions.
It is also applicable to study vortex phase diagrams of other conventional and unconventional type-II superconductors.
Thus, it is envisaged that the EPFC method becomes a general and powerful technique to investigate vortex physics.


\section*{Acknowledgements}
We acknowledge H. Takatsu for his contributions in crystal growth and R. Ikeda for helpful discussion.
This work was supported by the ``Topological Quantum Phenomena'' (Nos.~22103002 \& 22103004) Grant-in Aid for Scientific Research on Innovative Areas from the Ministry of Education, Culture, Sports, Science and Technology (MEXT) of Japan,
and by Grant-in-Aid for Scientific Research~(KAKENHI 26287078) from the Japan Society for Promotion of Science~(JPSJ)

\bibliography{../bib/Sr2RuO4,../bib/quenchedvortex,%
}

\end{document}